\documentclass[12pt,preprint]{aastex}
\usepackage{url,graphicx}
\usepackage[colorlinks]{hyperref}
\usepackage{amsmath,amsfonts,amssymb,amsthm}
\usepackage{listings}
\usepackage{enumitem}
\setlist[itemize]{noitemsep, topsep=-5pt}

\begin{document}

\title{Fisher Matrix for Beginners}

\author{David Wittman\\
  Department of Physics and Astronomy,  University of California, Davis}


\begin{abstract} 
  The Fisher information matrix is used widely in astronomy
  (and presumably other fields) to forecast the precision of future
  experiments while they are still in the design phase.  Although many
  sources describe the mathematics of the formalism, few sources offer
  simple examples to help the beginner.  This pedagogical document
  works through a few simple examples to develop conceptual
  understanding of the applications.
\end{abstract}

\section{Hot Dogs and Buns}

Consider a universe with two kinds of particles: hot dogs and buns.
Our model of physics is that hot dogs and buns are generally produced
in pairs, but that hot dogs occasionally are produced alone in a
different process. We want to know the pair production rate and the
hot-dog-only production rate.  The only measurements we can do are
counting hot dogs and buns in a given volume of space.

In this example, there are two {\it observables}: number of hot dogs
$n_h$ and number of buns $n_b$.  Each observable has some measurement
uncertainty, $\sigma_h$ and $\sigma_b$ respectively.  There are two
{\it model parameters}: pair production rate (call it $\alpha$) and
hot-dog-only rate ($\beta$).  We can write down the model as:
$$ n_h = \alpha + \beta $$
$$ n_b = \alpha $$ 
assuming that we survey a unit volume of space and
a unit time.

The whole point of the Fisher matrix formalism is to predict how well
the experiment will be able to constrain the model parameters, {\it
  before doing the experiment} and in fact without even simulating the
experiment in any detail.  We can then forecast the results of
different experiments\footnote{In this simplified example, different
  experiments could only mean larger or smaller surveys, which would
  change the size of the measurement errors.  But we will soon come to a more interesting example.} and look at tradeoffs
such as precision versus cost.  In other words, we can engage in {\it
  experimental design.}

This example is so simple that we can use our intuition to predict
what the Fisher matrix will predict.  When we get the data, we will
probably infer the pair-production rate from the number of observed
buns, and infer the hot-dog-only rate by subtracting the number of
observed buns from the number of observed hot dogs.  If our experiment
happens to count too many\footnote{By this I mean that the volume 
  surveyed randomly happens to contain more buns than the universal average.}
buns, it would not only boost our estimate of the pair production
rate, but would {\it also} depress our estimate of the hot-dog-only
rate.  So there is a {\it covariance} between our estimates of the two
parameters.  We can also see that the variance in our estimate of the
pair-production rate will be equal to (apart from some scaling factors
like the total volume surveyed) the variance in bun counts, but the
variance in our estimate of the hot-dog-only rate will be equal to
(again neglecting the same scaling factors) the {\it sum} of the
variances of the bun and hot dog counts (because of simple propagation
of errors).


The beauty of the Fisher matrix approach is that there is a simple
prescription for setting up the Fisher matrix {\it knowing only your
  model and your measurement uncertainties}; and that under certain
standard assumptions, the Fisher matrix is the inverse of the
covariance matrix.  So all you have to do is set up the Fisher matrix
and then invert it to obtain the covariance matrix (that is, the
uncertainties on your model parameters).  You do not even have to
decide how you would analyze the data!  Technically, the Fisher
information is an \textit{upper limit} on how precise an unbiased
estimator can be (i.e. its inverse is a lower limit on the variance of
the estimator); this is known as the Cram\'{e}r-Rao bound.

The fact that this approach involves no data analysis is a
double-edged sword. It makes forecasting easy, so you can swiftly
compare a variety of experimental designs.  But there is no guarantee
that you will ever create a data analysis method that reaches this
bound. Once you have settled on an experimental design, you are
advised to generate and analyze mock data to confirm that you can come
close to that bound, before committing resources to actually running
the experiment.

Here's the prescription for the elements of the Fisher matrix
$\mathcal{F}$. For $N$ model parameters $p_1, p_2,...p_N$, 
$\mathcal{F}$ is an $N\times N$ symmetric matrix.
Each element involves a sum over the observables.  Let there be $B$
observables $f_1,f_2...f_B$, each one related to the model parameters by
some equation $f_b = f_b(p_1,p_2...p_N)$.  Then
the elements of the Fisher matrix are
$$ \mathcal{F}_{ij} = \sum_b {1 \over \sigma_b^2} {\partial f_b \over
  \partial p_i} {\partial f_b \over \partial p_j}$$
assuming Gaussian
  errors on each observable, characterized by $\sigma_b$.
In this case, identifying $\alpha$ as $p_1$, $\beta$ as
$p_2$, $n_h$ as $f_1$ and $n_b$ as $f_2$, we find that\footnote{The student is strongly encouraged to work through this example in detail.}
$$\mathcal{F} = \left[ \begin{array}{cc}
{1\over \sigma_h^2} + {1\over \sigma_b^2} & {1\over \sigma_h^2} \\
{1\over \sigma_h^2} & {1\over \sigma_h^2}
\end{array} \right]$$

Inverting the 2x2 matrix yields the covariance matrix
$$\left[ \begin{array}{cc}
\sigma_b^2 & -\sigma_b^2 \\
-\sigma_b^2 & \sigma_b^2 + \sigma_h^2
\end{array} \right]$$
much like we expected.\footnote{Specifically, the following properties
  meet my expectations: the constraint on the pair production
  rate depends only on the bun measurement; the constraint on the
  hot-dog-only rate depends on both measurements; and the off-diagonal
  term is negative because a fluctuation in the hot dog rate induces an
  opposite-sign fluctuation in the pair-production rate.}  This
example is underwhelming because it was so simple, but even in this
case we have accomplished something.  The simple approach to data
analysis that we sketched above would yield the same covariances; and
we know the Fisher matrix result is the best that can be achieved, so
we can now be confident that our data analysis plan is actually the
best that can be done.

The full power is really evident when you consider cases with just a
few more observables and just a few more parameters.  It would be
extremely tedious to manually write out, say, a 4x4 matrix (for four
model parameters), each element of which is the sum of say 5 terms
(for 5 observables), and invert it.  But doing it {\it numerically} is
extremely easy; basically, a few lines of code for taking the
derivatives, wrapped inside three nested loops (over Fisher matrix
columns and rows and over observables), plus a call to a matrix
library to do the inversion.  For that small amount of work, you can
forecast the (maximum possible) efficacy of an extremely complicated
experiment!

\section{Fitting a Line to Data}

As a second example, consider fitting a straight line to some data: $f
= ax+b$.  Imagine that you can afford to take only two data
points; at what values of $x$ would you choose to measure?
Intuitively, we would say as far apart as possible, to obtain the best
constraint on the slope.  With the Fisher matrix, we can make this
more quantitative. (Again, note that the Fisher information matrix
approach does not tell you {\it how} to fit a line, or in general how
to analyze your data.)

In this case, our two observables are {\it not} qualitatively
different, like hot dogs and buns.  They are simply measuring the same
kind of thing at two different values of $x$.  But they can
nonetheless be considered two different observables united by a common
model: $f_1 = ax_1 + b$ and $f_2 = ax_2 + b$.  The Fisher matrix is
then\footnote{Again, the student is strongly encouraged to work this
  through!}
$$\mathcal{F} = \left[ \begin{array}{cc}
{x_1^2\over \sigma_1^2} + {x_2^2\over \sigma_2^2} & {x_1\over \sigma_1^2} +{x_2\over \sigma_2^2}\\
{x_1\over \sigma_1^2} +{x_2\over \sigma_2^2} & {1\over \sigma_1^2} +{1\over \sigma_2^2}
\end{array} \right]$$

Inverting this and simplifying with some slightly tedious algebra, we
obtain the covariance matrix
$$ {1\over (x_1-x_2)^2} \left[ \begin{array}{cc}
\sigma_1^2 + \sigma_2^2 & -x_1\sigma_2^2 - x_2\sigma_1^2 \\
 -x_1\sigma_2^2 - x_2\sigma_1^2 &  x_1^2\sigma_2^2 + x_2^2\sigma_1^2
\end{array} \right]$$

In other words, the variance on the slope is ${\sigma_1^2 + \sigma_2^2
  \over (x_1-x_2)^2}$, which makes perfect sense because it's the
variance in ${y_2-y_1\over x_2-x_1}$.  The other elements are somewhat
more complicated, such that you would not have guessed them without
grinding through the least-squares fitting formulae.  In fact, we can
gain new (at least to me) insight by looking at the covariance between
slope and intercept: because the numerator contains odd powers of $x$,
we can make it vanish!  Specifically, if we choose ${x_1\over x_2} = -
{\sigma_1^2 \over \sigma_2^2}$, we completely erase the covariance
between slope and intercept.\footnote{The covariance matrix can also
  be diagonalized without changing $x_1$ or $x_2$, by rewriting $f$ as
  a function of $x-x_0$ and carefully choosing $x_0$; in other words,
  by generalizing the concept of the ``intercept'' of the
  function. Thanks to Zhilei Xu and Duncan Watts for pointing this
  out.}  If this were an important consideration for your experiment,
you'd be glad for the insight.

More commonly, though, we'd have more than two data points.  If we can
afford a third, should we put it in the middle or make it as extreme
as possible as well?  Answering this question analytically would be
extremely tedious, so now is the time to code it up.

\textbf{Exercise:} Write a code to generate and invert the Fisher
matrix for fitting a line $y(x)$ to $n$ data points. The input should
be a vector of $x$ values (where the $y$'s are measured) and a vector
of measurement uncertainties, $\sigma$, for the $y$'s.  Note that the
$y$'s themselves are not needed, and in this basic line-fitting
example there is no uncertainty associated with the $x$'s.  Your code
must allow for $\sigma$ to be a different value for each data point
(`heteroscedastic data') even if we don't use that feature right now.

Running this code with $x=(-1,1)$ and $\sigma=(0.1,0.1)$, I get:
\begin{verbatim}
[[ 0.005  0.   ]
 [ 0.     0.005]]
\end{verbatim}
which confirms the results (or confirms the script, depending on how
you look at it).  To further test the script, add a third point (with
the same $\sigma$) at $x=0$; this should not improve the slope
constraint, but should help the intercept.  The result is:
\begin{verbatim}
[[ 0.005       0.        ]
 [ 0.          0.00333333]]
\end{verbatim}
This confirms that we are correctly interpreting the order of the
matrix elements that numpy spits out.  Finally, move the third point
to $x=1$ (imagine that this is at the extreme end of where you can
measure).  The result is:
\begin{verbatim}
[[ 0.00375 -0.00125]
 [-0.00125  0.00375]]
\end{verbatim}
This helped constrain the slope {\it and} the intercept, at the cost
of some covariance.

{\bf Exercise:} (a) Use your code to generate the Fisher matrix for fitting a line to {\it
  one} data point and attempt to invert it to obtain the covariance
matrix. What happens and why? Explain why infinite covariance does not
necessarily imply zero information. (b) If we now take a second data
point at the {\it same value of x}, compare what happens to the
information and the covariance.

\section{Plotting your forecast}

Presenting a covariance matrix as a table is not very useful in real
life. You presumably want to compare different experiments, and your
audience is not going to see the impact of all those numbers in table
form. Let's draw our covariance matrix as an ellipse in parameter
space; it will then be easy to draw multiple ellipses on the same
plot. I'll use Python and matplotlib for my implementation.  Even if
your model has more than two parameters, we will only ever plot the
covariance for two parameters at a time.

Any covariance
  matrix has the form
\begin{equation}
  \label{eq-covmatrix}
  {C} \equiv \begin{bmatrix}
\sigma_x^2 & \sigma_{xy}^2 \\
\sigma_{xy}^2 & \sigma_y^2 
\end{bmatrix} 
\end{equation}
which is symmetric because $\sigma^2_{xy}=\sigma^2_{yx}$
always. Furthermore, because
$(\sigma^2_{xy})^2 \le \sigma^2_{x}\sigma^2_{y}$ (the covariance can't
be larger than the variance), the determinant of this matrix is always
positive; both eigenvalues are real and positive; and the two
eigenvectors are perpendicular.  These properties map onto an ellipse:
the eigenvalues represent (the squares of) the semimajor and semiminor
axis lengths of the ellipse, and the eigenvectors represent the
directions of the principal axes.\footnote{In the limit where
  $(\sigma^2_{xy})^2 = \sigma^2_{x}\sigma^2_{y}$ the ratio of
  eigenvalues is infinite and the determinant vanishes, but you can
  still think of this as an ``ellipse'' with a minor axis length of
  zero.}  I'm not going to prove that, but you should check that it
makes some sense: an ellipse has three free parameters (major and
minor axes and rotation angle) as does a symmetric $2\times2$ matrix;
and axis lengths could only be represented by positive real
eigenvalues.  A general $2\times2$ matrix has four free parameters and
may not have real eigenvalues so could not possibly represent an
ellipse; and a diagonal $2\times2$ matrix has only two free
parameters, so at best (if it happens to have positive numbers in the
diagonal entries) it could represent an ellipse with zero rotation.

For a 2x2 symmetric matrix, there's a (reasonably) simple closed-form
expression for the eigenvalues:
\begin{equation}
  \label{eq-eigenvals}
  \lambda_{1,2} = \frac{1}{2}(\sigma_x^2+\sigma_y^2 \pm\sqrt{(\sigma_x^2-\sigma_y^2)^2+4 \sigma_{xy}^4}\,\,)
\end{equation}
The square roots of these eigenvalues are the semiaxis lengths of your 68\% confidence
ellipse---apart from a scaling factor I will now explain. Let's
look at a covariance matrix where we can read the eigenvalues
right off:
\begin{equation}
\begin{bmatrix}
2 & 0 \\
0  & 1
\end{bmatrix} 
\end{equation}
Clearly, the eigenvalues are 2 and 1. The variances in the parameter estimates will be 2 and 1, respectively, so the
uncertainties will be $\sqrt{2}$ and $\sqrt{1}$.  The zero
off-diagonal elements tell us there is no correlation between the
parameters, so you would think the confidence ellipse would be plotted
with a horizontal semimajor axis of $\sqrt{2}$ and a vertical
semimajor axis of $\sqrt{1}$.   But there is an important subtlety
when going from confidence
intervals (CI) on one parameter at a time to the \textit{joint
  confidence interval} on two parameters, as follows.

\begin{figure}[h]
  \centering
  \includegraphics[width=0.6\textwidth]{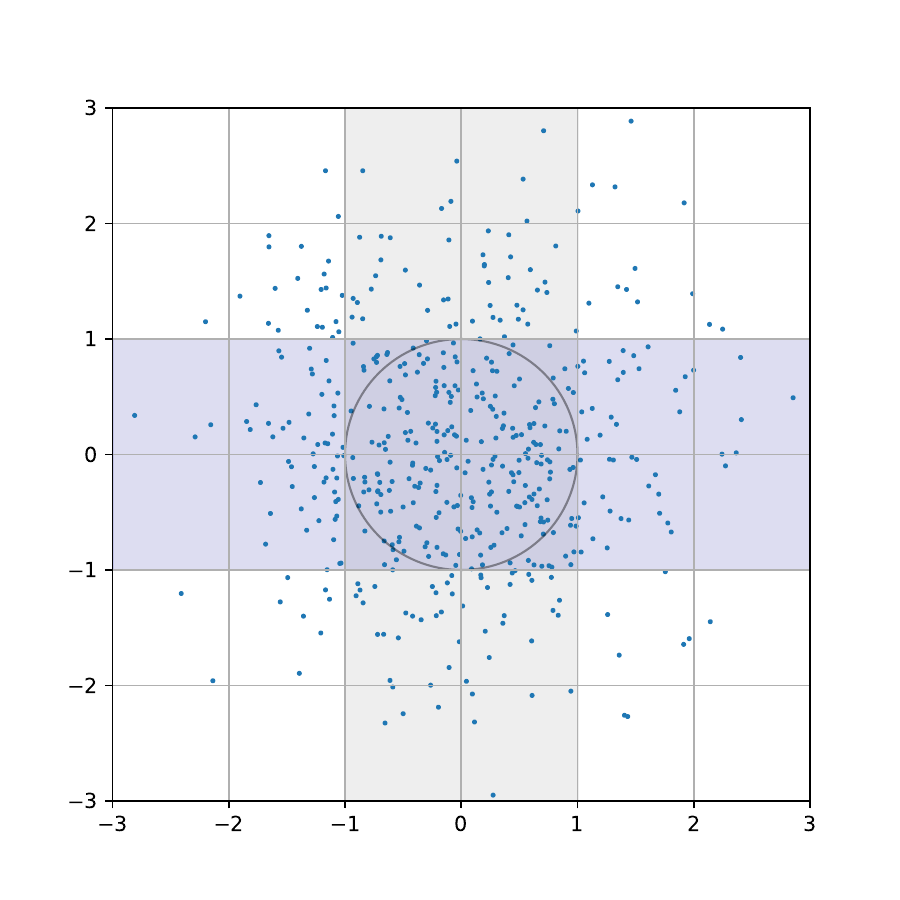}
  \caption{A distribution that is standard normal in both $x$ and
    $y$. The shaded rectangles highlight how 68\% of the points have
    $x$ in [-1,1] and 68\% have $y$ in [-1,1]. But the circle
    demonstrates that {\it far fewer} are within 1 unit of the
    origin. Therefore, the \textit{joint} 68\% CI should be
    larger---by a factor of 1.52 according to a 2-D Gaussian integral.}
  \label{fig-2dstandardnorm}
\end{figure}

To most clearly see the 1-D vs 2-D distinction, let's rescale the units of our
``x'' parameter so our covariance matrix is simply
\begin{equation}
\begin{bmatrix}
1 & 0 \\
0  & 1
\end{bmatrix} 
\end{equation}
You already know this implies a 68\% CI of $\pm1$ on each of parameter
individually. But the \textit{joint CI cannot be a circle of radius
  1}. Figure~\ref{fig-2dstandardnorm} illustrates why, with a cloud of
points illustrating the likelihood distribution corresponding to this
covariance matrix. \textit{The unit circle encloses less than 68\% of
  the points.} This is because of dimensionality: we were given that
the vertical gray stripe includes 68\% of the points (because the 68\%
CI is $\pm1$ on this parameter), so the unit circle, which has the
same width but covers much less area, must contain a substantially
smaller fraction of the points. If you do the 2-D Gaussian integral to
find the fraction of the points (ie the fraction of the likelihood)
within a given radius, you will find that 68\% lie within a radius of
1.5158. Therefore, you must draw your joint (68\%) CI as an ellipse
with principal semiaxes 1.5158 times the (square roots of the)
eigenvalues of your covariance matrix.\footnote{If you want a joint
  $2\sigma$ (95.4\% CI), use 2.2977 times the eigenvalues, and for
  $3\sigma$ (99.7\% CI), use a factor of 3.5292. These values all come
  from the same 2-D Gaussian integral, just with different limits.}
The dimensionality argument and the factor of 1.5158 apply to
\textit{any} ellipse describing a 2-D joint CI, no matter its axis
ratio or rotation.\footnote{To see this, imagine that you can always
  reparametrize to obtain uncorrelated parameters---thus making an
  ellipse aligned with the principal axes of your parameter
  space---and then rescale the units of your parameters to obtain unit
  variance.}

\textit{Rotation:} I mentioned above that the covariance matrix
eigenvectors determine the directions of the ellipse principal axes,
but there is an efficient shortcut. For plotting purposes, you
simply want to know the position angle (PA)\index{position angle} of
the major axis. It takes a page or two of algebra to work this out
from the eigenvectors,\footnote{See, for example,
  \url{http://scipp.ucsc.edu/~haber/ph116A/diag2x2_11.pdf}.}  but the
result simplifies down to
\begin{equation}
  \label{eq-theta}
  \tan 2\theta = \frac{2 \sigma_{xy}^2}{\sigma_x^2-\sigma_y^2}
\end{equation}
where $\theta$ is measured counterclockwise from the $x$ axis. Note
that an ellipse is a `headless vector': the direction it points can
only be determined modulo 180$^\circ$. So a PA of, say, 35$^\circ$
would be the same as a PA of 215$^\circ$. The procedure described here
will give you a PA in the range $\pm90^\circ$.

A few other quick conceptual points:
\begin{itemize}
\item apply this knowledge to plots that you \textit{see}, not just
  plots that you make. If you see a forecast that plots a perfect
  ellipse, you can be pretty sure that it was made with the Fisher
  matrix. In contrast, a more detailed forecast would involve
  simulating data for the proposed experiment, then analyzing that
  mock data---and such an analysis would rarely come up with a perfect
  ellipse for its confidence intervals. The ellipse is a sign that the
  forecast does not account for the nitty gritty details of the
  experiment, and that it's the \textit{best} that could possibly be
  done if the experiment is run.
\item \textit{where} in parameter space do you center your forecast
  ellipse? The Fisher machinery is agnostic: if you didn't use the
  value of a particular model parameter at any point in your forecast,
  you can center the ellipse on \textit{any} value of that
  parameter. Most people would center it on the currently accepted
  best value, to better compare the forecast with current constraints.
\end{itemize}

\section{Fiducial models}

In the previous paragraph I wrote ``\textit{if} you didn't use the
value of a particular model parameter at any point in your
forecast...''.  In the hot dog/bun and line-fitting examples, we did
not use them. The forecast is the same \textit{regardless} of
what the actual slope and intercept turned out to be. This is
remarkable. It happened because the observables were \textit{linear}
in the parameters.  Because of that linearity, the derivatives you put
into $\mathcal{F}$ did not depend at all on the model parameters.

You will find many situations in which those derivatives \textit{do} depend on the
model parameters.  For example, consider $y=\exp(-x/x_0)$: $${\partial
  y\over \partial x_0} = {x\over x_0^2} \exp(-x/x_0).$$ Conceptually,
it makes sense that $x_0$ would figure into the precision of your
experimental constraints; after all, the larger the exponential scale
length, the less able you are to detect any decline if you are
sampling a fixed range of $x$.

Therefore, you must assume a value of the model parameter in order to
obtain a forecast.  Is this circular reasoning? No, you are merely
forecasting the precision of the experiment, not its outcome. If the
precision depends on the parameter values, you must make your guess
for what they are---you must declare your \textit{fiducial
  model}---and forecast based on that.  To do a thorough job, you should
repeat the forecast for some other parameter values and then assess
how much the precision really depends on those model parameters.

This is a good reminder that the whole Fisher matrix formalism is
based on how the data respond to an \textit{infinitesimal}
perturbation of the model parameters. Remember the derivatives
$\frac{\partial f_b}{\partial p_i}$? Those are infinitesimal
perturbations. It's a bit of an extrapolation to assume that the same
response is true for large perturbations. (Exception: for some
parameters, such as slope and intercept in the line-fitting case, we
can prove mathematcically that the extrapolation is exact.) This
statement cuts both ways: the forecast may be overly optimistic due to
this property, but it may also be overly pessimistic.  Here's a
striking example: imagine your model is a parabola $y=a(x-b)^2$, so
the derivative with respect to $b$ is $2a(x-b)$. Therefore, any data
taken at $x\approx b$ contributes (in this formalism) \textit{zero}
information about $b$. Graphically speaking, the nature of a parabola
is that an infinitesimal horizontal shift has, to first order,
\textit{no effect} near the vertex.  But \textit{finite} shifts will
have an effect; they are simply not captured by the Fisher formalism.
So the Fisher forecast dramatically underestimates how points near the
vertex can help constrain $b$.

\section{Priors}

A {\it prior} represents your knowledge of the model parameters prior
to the experiment.  In the context of forecasting constraints using
the Fisher matrix, it represents how {\it precise} your prior
knowledge is.  This is easy to visualize in terms of the covariance
matrix.  In the line-fitting example, imagine that by some revealed
knowledge (from a previously published experiment, perhaps, or because
of some theoretical consideration), you already know the slope to
within $\sigma_{\rm slope,prior}$ and the intercept to within
$\sigma_{\rm intercept,prior}$.  You could then represent your prior
knowledge with the covariance matrix

$$ \left[ \begin{array}{cc}
\sigma^2_{\rm slope,prior} & 0 \\
0 & \sigma^2_{\rm intercept,prior}
\end{array} \right]$$
(again, we are assuming Gaussian uncertainties).  You can invert this
into a Fisher matrix
$$ \left[ \begin{array}{cc}
\sigma^{-2}_{\rm slope,prior} & 0 \\
0 & \sigma^{-2}_{\rm intercept,prior}
\end{array} \right]$$
and add it to your experiment's Fisher matrix.  This represents the
total information available (the full name of the Fisher matrix is the
Fisher information matrix, after all; it is the inverse of the
variance).  Now invert that total matrix into a new covariance matrix
to yield a forecast of the covariances you will have after doing your
experiment.

That's a conceptual view.  In practice, you may have a prior on only
one of the parameters, and you cannot represent this with an
invertible matrix.  So in practice people just add $\sigma^{-2}_{\rm
  prior}$ to the appropriate diagonal element of $\mathcal{F}$.

It's fun to add a prior and watch your confidence intervals shrink. On
the flip side, you may find that your experiment yields so little
information on some parameter that the Fisher matrix is not even
invertible. Then you \textit{need} to add a prior to go further. This
alone provides a valuable insight into your proposed experiment.

Note that priors added through this formalism are necessarily
Gaussian.  You may want to apply, say, a ``tophat'' prior (uniform
from 0--1 and vanishing outside that range), and you can do that
easily at the data analysis stage, but this formalism does not allow
you to forecast that easily.  You will have to approximate it with a
Gaussian prior.

If you know a parameter very, very well, you will get an extremely
large number in that Fisher matrix element.  In the limit where you
want to fix a parameter completely, you may want to just remove it
(i.e., remove the corresponding column and row) from the Fisher matrix
rather than put a ginormous number there, to protect yourself from
numerical instabilities.\footnote{Numerical instabilities in inverting
  matrices are more common than beginners may think. If you find that
  your Fisher matrix has a large condition number (ratio of largest to
  smallest eigenvalue), you may try rescaling some parameters to
  bring the condition number down and insure numerical stability when
  inverting it.} 

\section{Nuisance Parameters and Marginalizing}

In the line-fitting example, what if you recorded a bunch of data and
derived a fit, but are interested {\it only} in the slope and not at
all in the intercept?  Then the intercept can be considered a {\it
  nuisance parameter}, and you would like to integrate or {\it
  marginalize} over possible values of this parameter when placing
confidence limits on the slope.

At least, that's the way of viewing it when you're actually analyzing
your data.  In terms of forecasting, the Fisher matrix makes this
trivial: the values in $\mathcal{F}^{-1}$ {\it are} the marginalized
variances.  This is repeating the point made by
Figure~\ref{fig-2dstandardnorm}: the joint CI takes some work to
derive from the individual CI. The latter can simply be read off the
covariance matrix.

More commonly, nuisance parameters are those which describe the
experiment and not the scientific model.  For example, you may have a
calibration in your measurement machinery which is uncertain by 10\%.
You should explicitly write a variable calibration factor into the
model and add expand the Fisher matrix to include this variable,
taking all the necessary derivatives.  Then, represent the 10\% prior
on the calibration parameter by adding to the appropriate diagonal
element of the Fisher matrix before inverting.

In some cases, you may find that the Fisher matrix forecasts a tight
constraint on your calibration parameter even {\it without} specifying
a prior on it.  If so, you have stumbled into a {\it self-calibration
  regime}.  The parameter space and the experiment might work together
so that you can use the data to simultaneously solve for model {\it
  and} calibration parameters, without incurring much extra
uncertainty on the model parameters.


{\bf Exercise:} Return to the exercise in which we forecast
constraints provided by fitting a line to one data point (or multiple
data points over a very limited range of the independent
variable). Imagine you have prior information about the intercept from
some other experiment or from theory.  Add this information to the
Fisher matrix and show that the covariance matrix no longer blows
up. Interpret this result.

{\bf Exercise:} (a) A typical galaxy model says that the intensity as
a function of radius is $I(r)=I_0 \exp(-{r\over r_0})$. At each value
of $r$ there is some uncertainty in your measurement of $I$; let us
call this $\sigma_I$. Find the covariance matrix if you measure $I$ at
three values of $r$: 0, $r_0$, and $2r_0$ (you may want to write a
script to do this). (b) In real images there is always background
light so you fit a model $I(r) = I_0 \exp(-{r\over r_0}) + b$ where $b$
is a uniform background. Find the covariance matrix for estimating all
three parameters from the data. (c) For physical reasons, the amount
of background light cannot be arbitrarily large, nor can it be less
than zero. Put some prior on $b$ and see how that affects your
constraint on $I_0$ and $r_0$. Note that assigning this prior is a bit
of an art and is quite distinct from assigning $\sigma_I$, which
should be known rigorously based on the properties of your camera.

\section{Multiple Experiments}
What if we have multiple experiments that constrain a model?  The
Fisher matrix makes it easy to forecast the precision of a joint
analysis: just add the Fisher matrices of the experiments, and invert
the summed matrix.  To see this, just consider the different
experiments as different observables; we were already summing over the
$B$ observables of a given experiment to produce each element of
$\mathcal{F}$, so now we simply sum over the $B_1$ observables of the
first experiment, the $B_2$ observables of the second experiment, and
so on.  Note that the different experiments' observables do not at all
have to depend on the model in the same way!  The ${\partial f_b \over
  \partial p_i}$ terms represent each particular observable's
relationship to the model.

If you are combining multiple experiments and each experiment has
multiple nuisance parameters, $\mathcal{F}$ can get very large (in
terms of number of elements $N_{\rm param}^2$, not the size of the
elements).  If you need to reduce the size of $\mathcal{F}$ to invert
it, there may be a workaround. {\it If} the nuisance parameters of
each experiment are uncorrelated, you can find the covariance matrix
for the first experiment, then remove the rows and columns
corresponding to the nuisance parameters, and invert to find the
``marginalized Fisher matrix'' for that experiment.  You then sum the
marginalized Fisher matrices for all the experiments and invert to get
the final covariance matrix.  Obviously you can't take this shortcut
for nuisance parameters which are correlated from experiment to
experiment.  We tend to assume that these correlations don't exist and
that therefore we can take the shortcut of marginalizing first, but
think carefully when you do this.  Don't just do it by default.

\section{Some asterisks}\label{sec-asterisks}

In some experiments the uncertainties on the data points will not be
Gaussian. The more general expression for the Fisher matrix elements
is:
$$ \mathcal{F}_{ij} = \frac{\partial \ln \mathcal{L}}{\partial
  p_i}\frac{\partial \ln \mathcal{L}}{\partial p_j}$$ where
$\mathcal{L}$ is the likelihood and the partials, as noted above, are
evaluated at the fiducial model values.

In many sources, you will see the Fisher matrix defined with second
partials rather than with products of first partials.  According to
Wikipedia, the product of first partials is baked into the definition,
while the expression in terms of second partials follows if ``certain
regularity conditions'' are met.  This is related to another confusing
issue you will discover as you read more: there is a conceptual
distinction between the ``theoretical'' information matrix which we
have discussed here, and the ``observed'' information matrix which is
related to the curvature (second partials) of the likelihood surface
which you calculate from your actual observed data. The fact that
these two will equal each other under certain assumptions is what
allows you to make the forecast.


\section{Further reading}

These resources are biased toward the astrophysics literature because
that is what I am familiar with. Additional suggestions would be appreciated.

The Dark Energy Task Force report
(\url{http://arxiv.org/ftp/astro-ph/papers/0609/0609591.pdf}) has a
nice summary of the math behind Fisher matrix analysis starting on
p. 94.  The Findings of the Joint Dark Energy Mission Figure of Merit
Science Working Group
(\url{http://arxiv.org/PS_cache/arxiv/pdf/0901/0901.0721v1.pdf},
starting on p. 4) repeats much of this but also adds some cautionary
notes on numerical instabilities to watch for when doing the matrix
operations.

Two papers are often credited with
introducing Fisher forecasts to astrophysics:
\href{https://ui.adsabs.harvard.edu/abs/1997PhRvL..79.3806T/abstract}{Tegmark
  1997, {\it PRL} 79, 3806} and
\href{https://ui.adsabs.harvard.edu/abs/1997ApJ...480...22T/abstract}{Tegmark,
  Taylor \& Heavens 1997, {\it ApJ} 480, 22}.

Dan Coe's Fisher Matrices and Confidence Ellipses: A Quick-Start Guide
and Software
(\url{http://arxiv.org/PS_cache/arxiv/pdf/0906/0906.4123v1.pdf}) helps
you convert Fisher matrices to confidence contours and includes links
to relevant software packages.

Alan Heavens et al have developed generalized Fisher matrices, which allow
forecasts in more complicated cases (such as errors in both ``x'' and
``y'' directions, and also including other thorny situations.) See \url{https://ui.adsabs.harvard.edu/abs/2014MNRAS.445.1687H/abstract}.



\section*{Acknowledgements}

Helpful feedback and typo-catching for the original version of this
document was provided by Will Dawson and Andrew Bradshaw at UC Davis,
Zhilei Xu and Duncan Watts at Johns Hopkins, William Peria at the Fred
Hutchinson Cancer Research Center, Michael Saginaw somewhere out there
on the Web, and Ioana Zelko at Harvard University. I thank Jackson
Taylor (at West Virginia University) and Denise Schmitz (then at
Caltech) for encouraging me to submit this guide to arXiv, and Jackson
Taylor also for proofreading the expanded version submitted to arXiv.

\section*{Appendix: ellipse plotting in Python}
In Python, you can draw the forecast ellipse with this code:
\begin{lstlisting}
import numpy as np
import matplotlib.pyplot as plt
from matplotlib.patches import Ellipse

C = np.array([[ 0.00375,-0.00125], [-0.00125, 0.00375]])
eigenvals, eigenvecs = np.linalg.eig(np.mat(C))
major_axis = 2*1.52*np.sqrt(eigenvals[0])
minor_axis = 2*1.52*np.sqrt(eigenvals[1])
angle = np.degrees(0.5*np.arctan2(2*C[0,1],C[0,0]-C[1,1]))

myellipse = Ellipse(xy=[0,0],width=major_axis,height=minor_axis,angle=angle) 
ax = plt.subplot(111)
ax.add_artist(myellipse)

# plot decorations/details
plt.xlabel('slope') 
plt.ylabel('intercept')
plotsize = 0.2               # these three lines are needed
plt.xlim(-plotsize,plotsize) # because matplotlib is not
plt.ylim(-plotsize,plotsize) # auto-detecting the best plot size
ax.set_aspect('equal')
plt.show()
\end{lstlisting}

The result is:

\centerline{\includegraphics[width=0.7\textwidth]{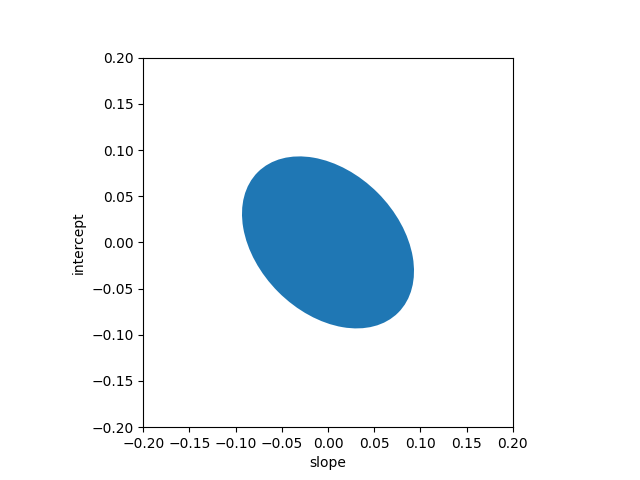}}

Important notes: 
\begin{itemize}
\item as explained in the main text, there is a factor of 1.52 in
  the axis lengths because we are plotting the \textit{joint} CI; and
  here there is an additional factor of 2 because matplotlib ellipses
  are specified in terms of their principal axes, rather than semi-axes.
\item this ellipse is centered at (0,0) but you will want to center it
  at your fiducial model.
\item when a matplotlib plot has no data, only drawings, you may need to tell
  matplotlib the limits on the plot, as is done here.
\item  matplotlib plots by default do not have equal aspect
  ratio. This can deceive the reader: the
  covariance matrix dictates that the ellipse should be at $-45$
  degrees, but it won't \textit{appear} to be at that angle if your
  plot is not square.  Pay attention to the plot aspect ratio to
  communicate clearly with your audience.
\end{itemize}

\end{document}